\pdfoutput=1

\documentclass{./sig-alternate-10pt}
\usepackage{type1cm}
\usepackage{graphicx}
\usepackage{xspace}
\usepackage{balance}
\usepackage{booktabs}
\usepackage[font={small}]{caption}
\usepackage{siunitx}
\usepackage{epsfig}
\usepackage{framed}
\usepackage{subfigure}
\usepackage{newfloat}
\usepackage{amsmath, amssymb, eurosym}
\usepackage{enumerate}
\usepackage{enumitem}
\usepackage[vlined,linesnumbered,ruled,noend]{algorithm2e}
\usepackage{lscape}
\usepackage{subfigure}
\usepackage{array}
\usepackage{pifont}
\usepackage{color, colortbl}
\usepackage{alltt}
\usepackage{listings}
\usepackage{multirow}

\def\addauthorsection{\ifnum\originalaucount>6
	\section{Additional Authors}\the\addauthors
	\fi}

\definecolor{Gray}{gray}{0.90}
\newcolumntype{a}{>{\columncolor{Gray}} p{1.7cm}}

\def\ttfntsize{9}
\makeatletter
\let\oldtexttt\texttt
\let\texttt\@undefined
\makeatother
\newcommand{\texttt}[1]{\fontsize{\ttfntsize}{\ttfntsize}\oldtexttt{#1}}
\makeatletter
\let\oldtt\tt
\let\tt\@undefined
\makeatother
\newcommand{\tt}{\fontsize{\ttfntsize}{\ttfntsize}\oldtt} 
 
\begin{document}

\title{The coin that never sleeps. The privacy preserving usage
of Bitcoin in a longitudinal analysis as a speculative asset.}

\numberofauthors{5}
\author{%
	\alignauthor{Emmanouil Karampinakis\\
		\affaddr{FORTH-ICS, Greece}\\
		\email{mkarabin@ics.forth.gr}}\\
	\alignauthor{Michalis Pachilakis\\
		\affaddr{FORTH-ICS, Greece}\\
		\email{mipach@ics.forth.gr}}\\
	\alignauthor{Panagiotis Papadopoulos\\
		\affaddr{FORTH-ICS, Greece}\\
		\email{panpap@ics.forth.gr}}
	\and 
	\alignauthor{Antonis Krithinakis\\
		\affaddr{FORTH-ICS, Greece}\\
		\email{krithin@ics.forth.gr}}\\
	\alignauthor{Evangelos P. Markatos\\
		\affaddr{FORTH-ICS, Greece}\\
		\email{markatos@ics.forth.gr}}\\
}

\maketitle
\begin{abstract}
Bitcoin is the first and undoubtedly most successful cryptocurrency to date with 
a market capitalization of more than 100 billion dollars. Today, Bitcoin has more 
than 100,000 supporting merchants and more than 3 million active users.
Besides the trust it enjoys among people, Bitcoin lacks of a basic feature a substitute
currency must have: stability of value. Hence, although the use of Bitcoin as
a mean of payment is relative low, yet the wild ups and downs of its value
lure investors to use it as useful asset to yield a trading profit.

In this study, we explore this exact nature of Bitcoin aiming to shed light in
the newly emerged and rapid growing marketplace of cryptocurrencies and compare 
the investment landscape and patterns with the most popular traditional stock market of Dow Jones.
Our results show that most of Bitcoin addresses are used in the correct fashion to preserve security and privacy of the transactions
and that the 24/7 open market of Bitcoin is not affected by any political 
incidents of the offline world, in contrary with the traditional stock markets.
Also, it seems that there are specific longitudes that lead the cryptocurrency in terms of bulk of transactions, but there is not the same correlation with the volume of the coins being transferred.

\end{abstract}

\section{Introduction}
\label{sec:introduction}
Besides the penetration of credit cards in our everyday life, we are still far from being a cashless society.
Banknotes (i.e., cash) still constitutes the basic currency used as a payment method with only 
14\% of people\cite{cardAdoption} using electronic types of money. 
However, there are fundamental drawbacks of cash for every participating entity: (i) for the banks there is the risk of forgery,
and the cost of transport, (ii) for the government there is money laundering, and (iii) for the user there is the need of physical presence and the
risk of loss and theft.

It is only 10 years ago, since the technology of the public distributed ledger (or blockchain~\cite{bitcoin}) came along with the birth of the first Bitcoin.
Bitcoin is the first decentralized digital currency (or cryptocurrency)
which uses cryptography to verify each monetary transaction in a peer-to-peer fashion without requiring \emph{any} centralized administration 
or regulation (central bank, government, etc.). 
Since the birth of Bitcoin, numerous other cryptocurrencies have been created  (more than 1638 nowadays!~\cite{topCoins}). Yet, 
the most popular and the most expensive one is Bitcoin, mainly due to its proven security and its wide 
use in online marketplaces. Indeed, today more than 100,000 merchants and vendors accept Bitcoin as payment~\cite{merchantsAdoption} 
and according to a recent study~\cite{cambridgeStudy} there are between 2.9 million and 5.8 million unique active cryptocurrency users with the vast majority using Bitcoin.

However, besides its initial scope to become a substitute currency, Bitcoin misses an important requirement: stability of currency value. For example, according to~\cite{bitcoinAsset},
within the same month the value of Bitcoin experienced an average daily change of 2\% when on the other hand, the exchange rate between the euro and the U.S. dollar 
had an average daily change of less than 1\%. Hence, it becomes apparent that the wild up and down swings of its value turn Bitcoin to be a \emph{speculative}~\cite{investBitcoins} 
\emph{asset}, which people use to yield a trading profit (same as with gold, silver, etc.) rather than use for their everyday payments.

In this study, we aim to investigate exactly this part of Bitcoin and explore the investment landscape of Bitcoin. Our scope is conduct the first to 
our knowledge analysis of the Bitcoin as an asset that will shed light in the economics of the worldwide Bitcoin market and compare it with the most popular traditional 
stock market of Dow Jones in an attempt to understand 
the different investing patterns in the global and unregulated market of cryptocurrencies. To achieve this, we download the entire Bitcoin's blockchain and focused on transactions from 2013 to 2017. By building our own parser, we extract every transaction that is potentially 
associated with Bitcoin investment (coin purchase). In summary, in this study we make the following contributions:

\begin{enumerate}
\item We run a full Bitcoin node and going backwards we obtain a large 5 year long dataset from blockhain, which includes 285,000 blocks from 2013 to the end of 2017.

\item We build our own parser to connect block outputs and inputs in order to recover all Bitcoin transactions and public keys from the blockchain. 
As a results we create a dataset of 270 million Bitcoin transactions from 320 million public keys, ignoring any other content.

\item We evaluate the integrity and confidentiality of Bitcoin, regarding the awareness of its users about intended practices that do not compromise the security and privacy of their transactions.

\item We classify investing transactions and we analyze the patterns of such transactions. Finally we compare the behaviour of Bitcoin transactions with the investing 
transactions of the traditional stock market of Dow Jones. Results of our analysis show that the Bitcoin market turns to be the largest market of worldwide scale that 
is uninterruptedly open 24/7. In addition, we see that the volume of transactions in Bitcoin it non-correlated with the one of traditional stock markets.
\end{enumerate}

\section{Background}
\label{sec:background}
\noindent{\bf Cryptocurrency:}
Cryptocurrency is a digital currency whose core component is cryptography not only to (i) secure its transactions and (ii) verify the transfer of assets between untrusted parties~\cite{decentralization,decentralization2}, 
but also to (iii) control the creation of additional units as a mechanism against hyperinflation.

\noindent{\bf Blockchain:}
Blockchain is a publicly available, distributed and append-only ledger. Every new registration to the ledger is called a block and it is cryptographically 
linked to each previous block containing a cryptographic hash of that block, a timestamp and a record of transactions. The blockchain is managed by a 
peer-to-peer network where each node following the protocol have it's own copy of the ledger. In addition, there are special nodes called \emph{Miners} that 
confirm all transactions of a specific (10 minutes) time window and produce new block to the chain. Before creating each new block, the miner has to complete 
a \emph{Proof of work (PoW)}~\cite{pow}, which consists of a complicated mathematical puzzle. The difficulty of this PoW is adjusted so as to limit 
the rate at which new blocks can be generated by the network. When a block is added in the chain, it is very difficult to change, since it requires another 
PoW and there are thousands of other miners pushing to add their own block. In order to change a block one must also change all the blocks that are mined 
on top of it, so a transaction inside a block is more secure if it is deeper in the blockchain. The majority decision is driving the network, and since the 
majority of the computational power is honest, the transactions are secure and only valid blocks are added.
Every finalized PoW creates a reward of newly generated coins for the miner. 

\noindent{\bf Bitcoin:}
The first and the most successful cryptocurrency is Bitcoin~\cite{bitcoin}. Specifically, in 2013 Bitcoin made its first prominent steps, where 
its value soared more than 10-fold in two months: from 22 dollars in February to a record of 266 dollars in April!
This rapid growth of Bitcoin drew the interest of several initiatives
who created their own coins e.g, Litecoin, Dogecoin, Peercoin, Feathercoin, Zetacoin, etc. Each of these \emph{altcoins} is designed to provide different 
features e.g., increased anonymity, transaction speed, privacy, proof-of-stake, DNS resolution, mining speed.

However, Bitcoin has specific advantages over altcoins that allow it to rest on its laurels: (i) Bitcoin owns the vast majority of the user base, (ii) it is easier to access 
and has more merchants~\cite{BTmerchands}, exchanges~\cite{BTexchange}, hardware~\cite{BThardware}, and software~\cite{BTsoftware} to support it, (iii) it is backed 
by the biggest developer ecosystem, (iv) contrary to altcoins, it has obtained sufficient trust among people as an alternative to national currencies e.g., (in
Zimbabwe~\cite{BTzimbabwe} and Venezuela~\cite{BTVenezuela}), (v) the recent Segregated Witness~\cite{segWit} protocol upgrade that increased the block capacity, boosted Bitcoin's prestige, and proved that it can also adopt and partially fix its scalability problem~\cite{scalabilityProblem}. Considering all these advantages, in this paper we focus our analysis only on the ecosystem of Bitcoin as the most 
representative cryptocurrency.

\section{Data Analysis}
\label{sec:analysis}

\begin{table}[t]
\begin{tabular}{l|l}
{\bf Metric} & {\bf Amount} \\ \toprule
Volume of Blockchain snapshot & 141 GBytes \\
Period of Blockchain snapshot & 5 years (2013-2017) \\
Bitcoin transactions & 270 million \\
Public Bitcoin addresses & 340 million \\
Bitcoin transfers & 1.4 billion \\ \bottomrule
\end{tabular} 
\caption{Summary of dataset}
 \label{tbl:summary}
\end{table} 
 
To conduct our analysis, we used a simple blockchain client for the Bitcoin blockchain. With our client, we were able to participate in the blockchain and
run a full Bitcoin node at the end of 2017 and we gather the raw blockchain of 141 GBytes  dating from 2013 to 2017. 
Every transaction input in the blockchain is a reference to an output from a previous transaction, consisting of a hash and an index of that 
previous output. In order to disassemble the chain of blocks and extract the Bitcoin transactions, we developed our own Bitcoin parser which recursively matches 
the input of a transaction with the output of a previous one, using the corresponding transaction hash. Our transaction-oriented parser completely ignores any data possibly stored in the blockchain as a comment but even inside a fake address, since we label every address and transaction hash with a unique identifier. 
After extracting the transactions and public keys we created a dataset of 270 million transactions and 
340 million unique public addresses. 

However, a transaction can have multiple inputs and multiple outputs (e.g., changes, fees): in our transactions dataset, 
we see transactions that may have more than a thousand outputs. Thereby, as a next step we further process our data by demultiplexing 
each transaction to a single coin transfer, from, or to a public address. Thus, we construct a compact and uniform dataset, that can be easily analyzed in terms of distinct transactions or addresses. 
As a result, each transaction entry is simplified to a quadruple, which includes the following:

\begin{enumerate}
\item the transaction id (identifier for the transaction hash)
\item the public key id (identifier of the address)
\item the value transferred (specifying if sent or received)
\item the timestamp
\end{enumerate}

In our dataset, we have 1.4 billion distinct such simplified transactions (or Bitcoin transfers). 
Table~\ref{tbl:summary} presents in summary an overview of our dataset.

\begin{figure}
	\centering
	\begin{minipage}{0.23\textwidth}
		\centering
		\subfigure[Bitcoin transactions per day of week.]{
			\includegraphics[width=0.95\linewidth]{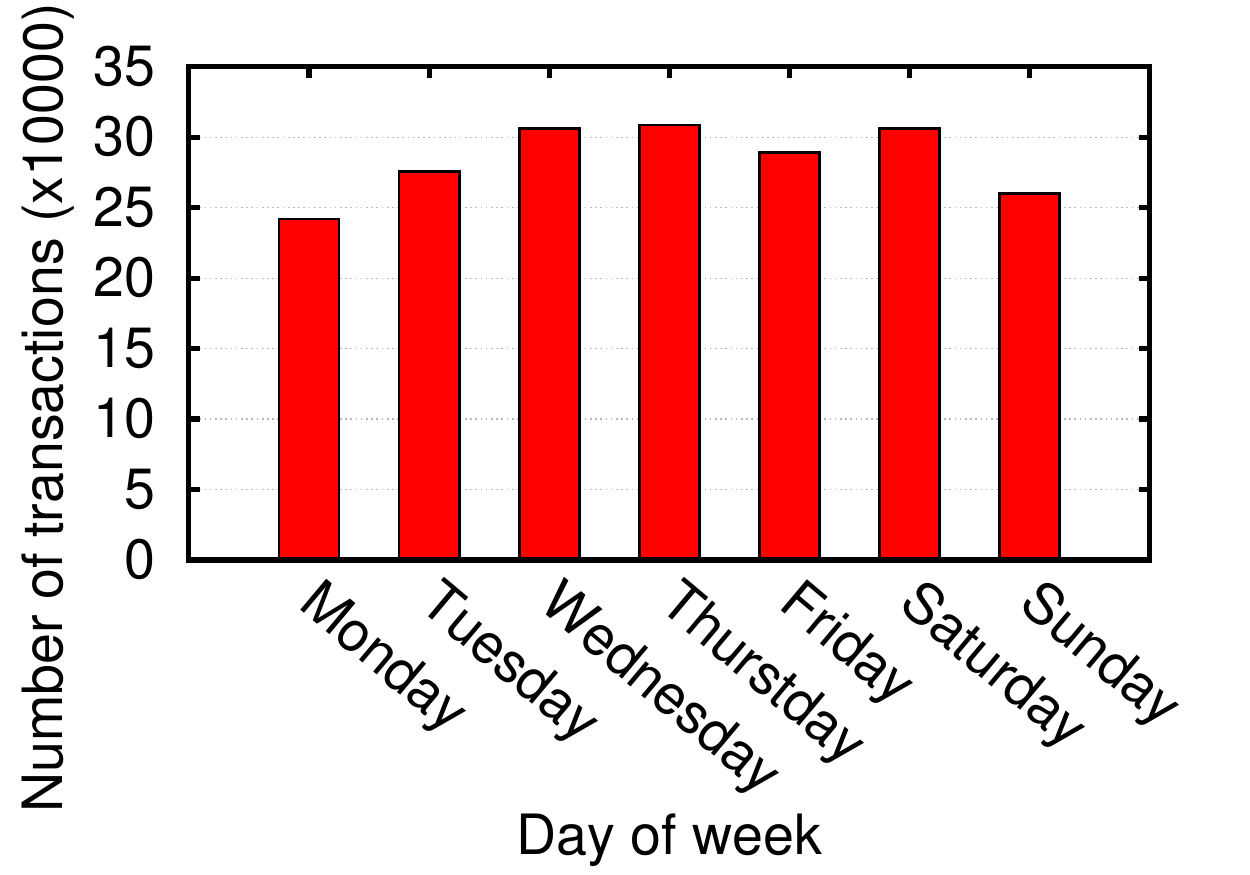}
			\label{fig:transactionsPerDay}
		}
	\end{minipage}
	\hfill
	\begin{minipage}{0.23\textwidth}
		\centering
		\subfigure[Bitcoins invested per day of week.]{
		\includegraphics[width=0.95\linewidth]{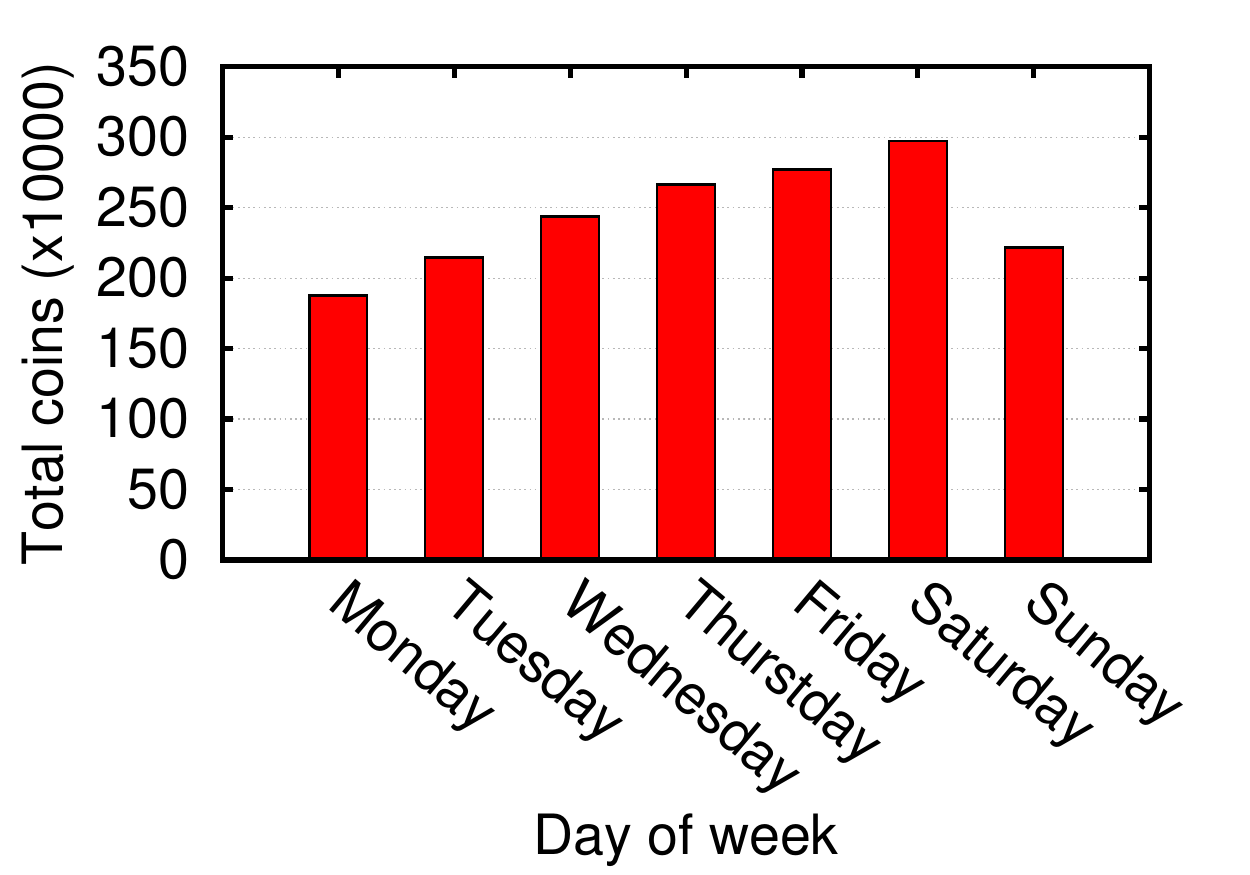}
		\label{fig:coinsPerDay}
	}
	\end{minipage}
	\label{fig:perDay}
	\caption{As we can see, the Bitcoin market contrary to traditional stock markets has no weekend brakes and transactions are  continuing.}
\end{figure}

\begin{figure}
	\begin{minipage}{0.23\textwidth}
		\centering
		\subfigure[Volume of transactions for the different hour of day.]{
			\includegraphics[width=0.95\linewidth]{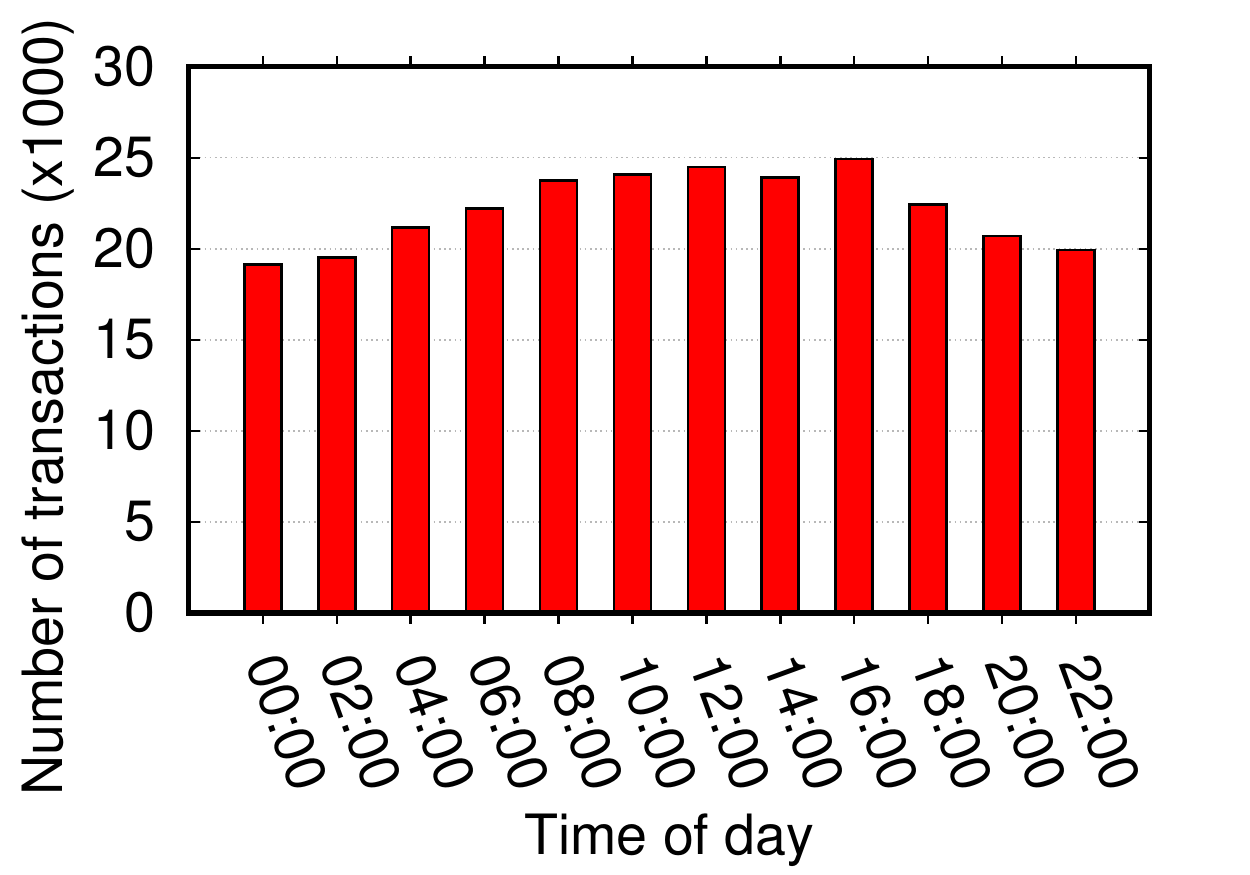}
			\label{fig:transactionsPerHour}
		}
	\end{minipage}
	\hfill
	\begin{minipage}{0.23\textwidth}
		\centering
		\subfigure[Bitcoins invested for the different hour of day.]{
			\includegraphics[width=0.95\linewidth]{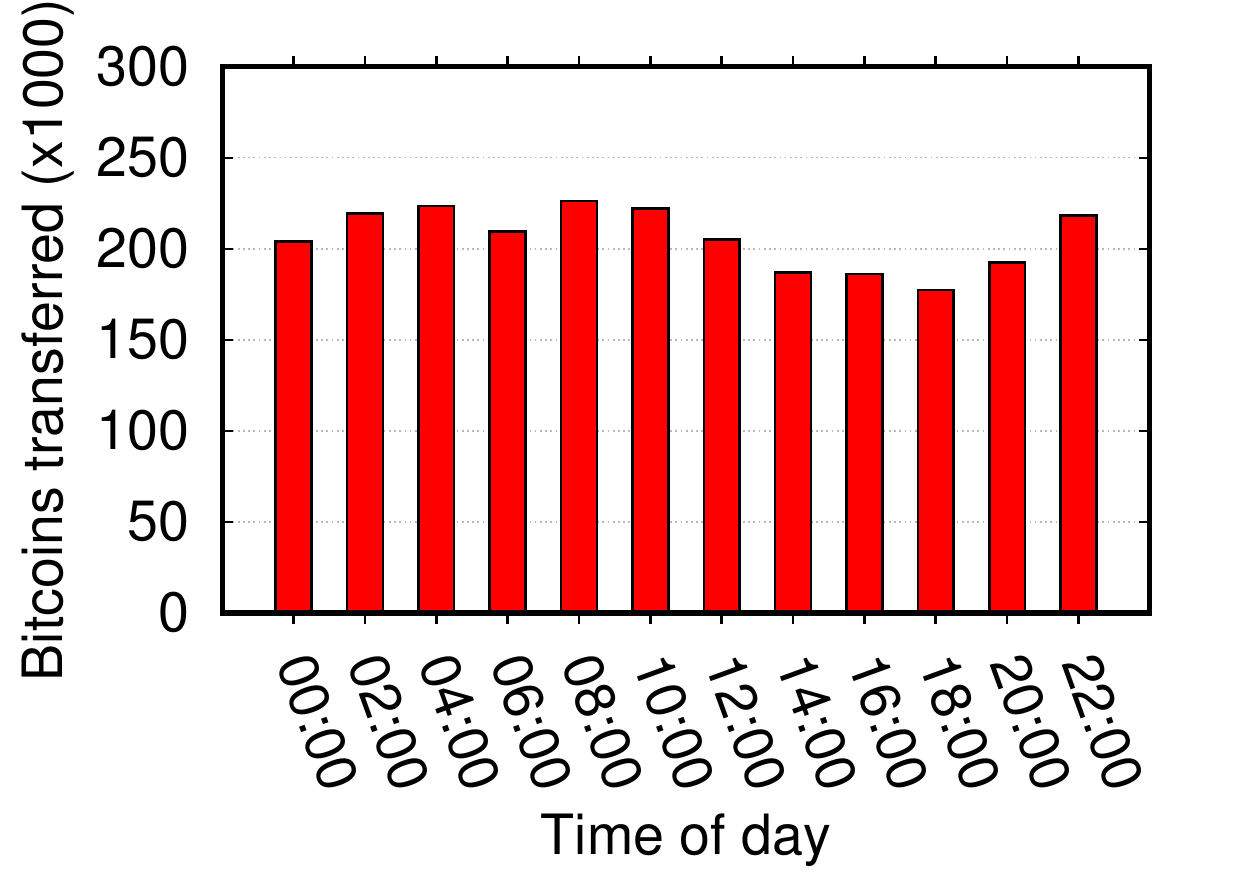}
			\label{fig:coinsPerHour}
		}
	\end{minipage}
	\label{fig:perhour}
	\caption{As we can see, there is a small diurnal cycle of Bitcoin market. However we clearly see that contrary to stock markets, the global market of Bitcoin continues to operate uninterruptedly across the entire day. EDT (Eastern Daylight Time) time zone was used.}
\end{figure}

\subsection{The Bitcoin ecosystem}
There is an major discrimination between traders and investors in any currency~\cite{investorsAndTraders}. Traders act fast and profit from the short-term price fluctuation, while investors hold on their investment, waiting for the perfect moment to sell, aiming to higher profits. In terms of Bitcoin, a trader will most likely take advantage of the high daily price fluctuation, and an investor who is indifferent to the short-term price volatility will not give up on the investment for many more days. We focus on the patterns that an investor is following in order to have a fair comparison with the traditional stock markets. Nevertheless, both are chasing profit, and our clustering on the patterns of investors provided indicative results on the notion that Bitcoin is highly used as a speculative asset.

As a first step, we wanted to evaluate the Bitcoin in terms of security and privacy. Bitcoin's integrity and privacy of transaction depends on the amount of honest nodes but also in the awareness of it's users about the intended practices. One major issue which we focused is the address reuse~\cite{reuse}. When addresses are re-used, they allow others to much more easily and reliably determine the user behind that address. Thus, we cluster the addresses regarding the number of transactions they appear and found out that the vast majority of addresses are only used once. We get that 86\% of all 342,744,669 addresses (2009-2017) are used only for 2 transactions (in one as an input and in one as output). Baring in mind that one user has multiple addresses, it is safe to presume that many, if not most of the users are well informed about the risks, and use the Bitcoin in a secure fashion.

\begin{figure}[t]
	\centering
	\includegraphics[width=0.85\linewidth]{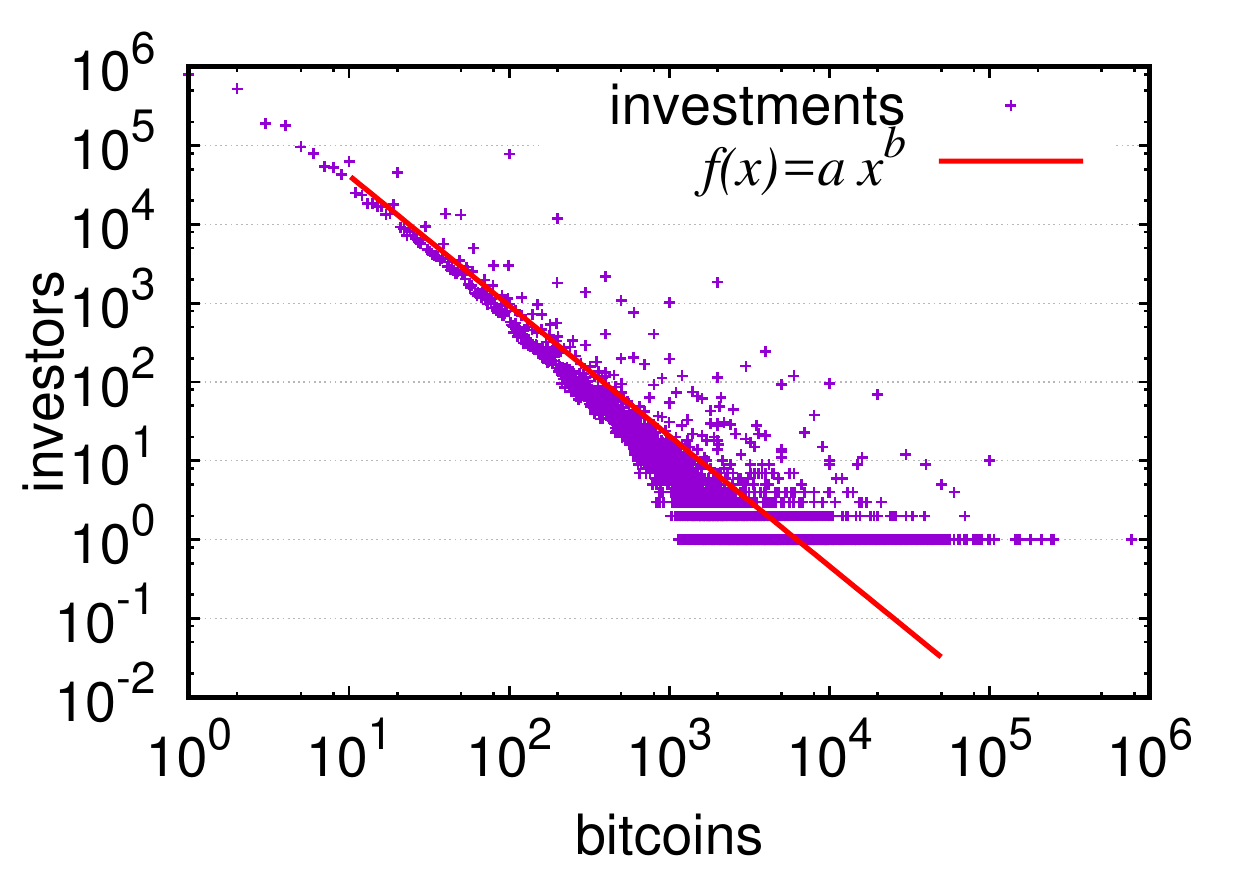}
	\caption{Investments in Bitcoin for the period 2013-2017. Power law distribution, with many users investing small amounts and few big players investing big amounts.}
	\label{fig:strategic_investments}
\end{figure}

In order to have a good idea of the investments in Bitcoin, we extracted all pubic addresses that behave as investors. We call an 
\emph{investor}, an address that for a period of an entire month, only receives coins with not a single outgoing transaction. 
We only take into consideration transactions that take place after January 1, 2013, when Bitcoin was a well known currency and started 
to have a significant price. 
As we can see in Figure~\ref{fig:strategic_investments} there are many addresses that invested small amounts in Bitcoin, but we observe 
an exponential decay on the investments for bigger values. Such a distribution is to be expected for an open currency like Bitcoin: 
many people act opportunistic over various times and convert money they can afford to cryptocurrencies. While there are a few big 
investors that seems to really believe in Bitcoin and bet huge amounts on it.

More precisely, we found more than 2 million Bitcoin addresses that seem to have invested in Bitcoin a value between 1 and 10 
coins, while the last 2 years of Bitcoin we observe a constant increase of investments. This is due to the popularity it has gained. As the leading cryptocurrency, many people see Bitcoin as a safe deposit for their money. All this trust 
people show to Bitcoin, sometimes results on strange patterns: people invest in Bitcoin not when its price is low, but when it starts 
rising, anticipating even higher price, thus starting a chain reaction of investments. Even the last months of 2017 where the price was 
extremely high and continuously rising there were more and more investments.
 
Next, we measure the daily and hourly patterns of the volumes of coins and transactions.
For these measurements, 3 periods of time were selected in 2017. One from January 2, until April 23, another from July 3, until August 13, and a final from September 4, 
until October 8. All those 27 weeks have something in common; they have really low average confirmation time on every block. This means that the timestamp of the block represents the actual time of the transaction accurately enough. For these weeks, we calculated the hourly volume of transactions and the coins transferred, converting the timestamps to EDT (Eastern Daylight Time), since New 
York is one of the most bitcoin-friendly cities with one of the biggest Stock Exchanges in the world.

As we can see in Figure~\ref{fig:transactionsPerDay} there is a significant variation on the confirmed number of transactions per day. On Monday, there are 242,000 transactions. Along with Sunday, those two days have the less transactions recorded throughout the week. In the middle of the week we observe many transactions. We must take into account that there is a time on Saturday (for every timezone in the globe), from which every Stock Exchange in the world is closed, until the first one opens on Monday morning. So any kind of investment and worldwide transaction is executed solely in Bitcoin.

The number of transactions is a good indicator of the total coins transferred every day of the week. As shown in Figure~\ref{fig:coinsPerDay} there is an interesting pattern, followed by the coins transferred. Monday is the day with the less coins in circulation, while as we move through the week there is an almost linear increase on the bulk of coins that is transferred every day. Saturday is the day with the highest amount of coins changing wallets, with almost 3 million Bitcoins being transferred. While on Sunday we observe a fall just like the previous Figure, and with fewer transactions we have fewer coins transferred.

Next, we set out to explore the possible patterns in finer granularity. Specifically, in Figure~\ref{fig:transactionsPerHour} we plot the volume of appeared transactions per 2 hour interval and as we can see, every noon on EDT there is a high volume of transactions. The hours with the most transactions recorded are 16:00 to 18:00. An interesting observation is that the New York Stock Exchange (NYSE) closes at 16:00. So, for the very first two hours after the closure of NYSE, we observe an increase, with 12,566 confirmed transactions. In general there are way more transactions taking place at noon, and much less around midnight and until the morning. This pattern seems to be representative and realistic for the selected time zone, indicating that USA and especially the East coast is dominating the transactions in the Bitcoin network.

We easily distinguished that the bulk of transactions is taking place at daytime of the EDT, but Figure~\ref{fig:coinsPerHour} does not seem to reveal the same for the amount of coins transferred. At noon of our time zone we observe the fewer coins per hour, while there are significant peaks at 04:00, at 09:00 and at 22:00. This is not something unexpected, it is just indicating that USA and Canada might be the leading countries in terms of transactions in Bitcoin, but there are bigger players around the world. We must not forget that Bitcoin exist where prosperity and technology exist, and apart from USA and Canada, there are many countries in Europe that embrace cryptocurrencies, and China which, as it seems, is the country with the highest mining power on Bitcoin.

\subsection{Bitcoin Vs. Stock Markets}

After exploring the characteristics of the Bitcoin market, in this section, we compare its behavior with the behavior we are used to see in 
traditional stock markets. Of course, the most representative such stock market is none other than the most popular one: Dow Jones. Hence, 
we gather publicly available data regarding the Dow Jones and Bitcoin for the period from January 2014 to March 2018. The data we collect 
include the volume (number of transactions) and value per day. Regarding the later, for Bitcoins we use its value in USD and for Dow Jones 
we use the Industrial Average (DJIA) stock market index. DJIA shows how 30 large publicly owned companies based in the United States have 
traded in the stock market during a standard trading session.

In the first experiment, we set out to explore the volume of transactions for both Bitcoin and Dow Jones and the pattern they follow across 
time. To have both in the same scale we conduct normalization using the following formula:
\[f(x) = \frac{x - min(distr)}{max(distr) - min(distr)}*100.0\%\]
\noindent where $x$ is the number of transactions, $min(distr)$ is the minimum value  and $max(distr)$ is the maximum 
value from the distribution.

\begin{figure}
	\centering
	\begin{minipage}{0.45\textwidth}
		\centering					
			\includegraphics[width=1.06\linewidth]{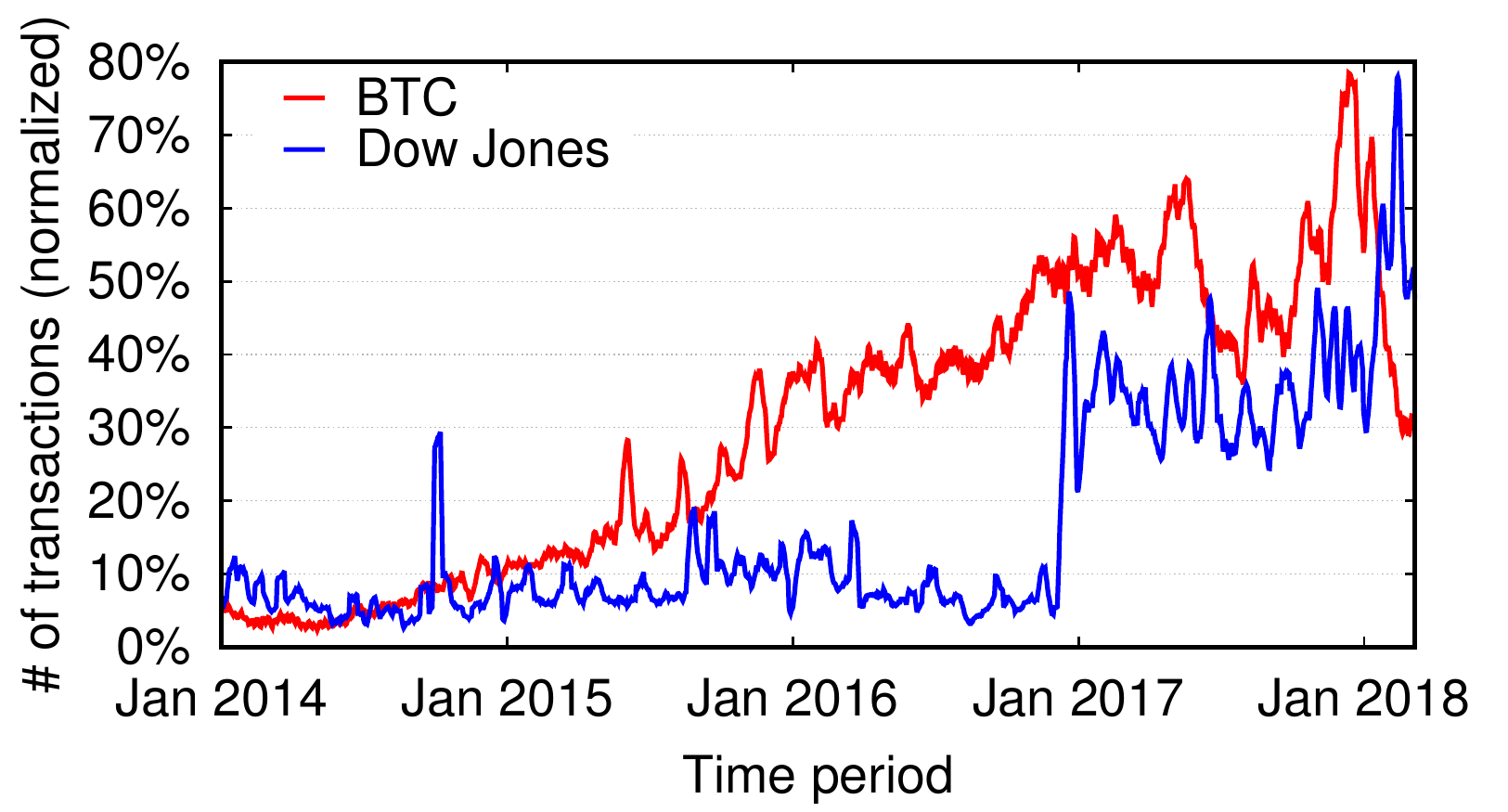}
			\caption{Number of transactions per day for Bitcoin and Dow Jones across the 4 years period. There is no clear correlation shown.}	
			\label{fig:volumeSince2014}	
	\end{minipage}
	\hfill
	\begin{minipage}{0.45\textwidth}
		\centering
		\includegraphics[width=1.06\linewidth]{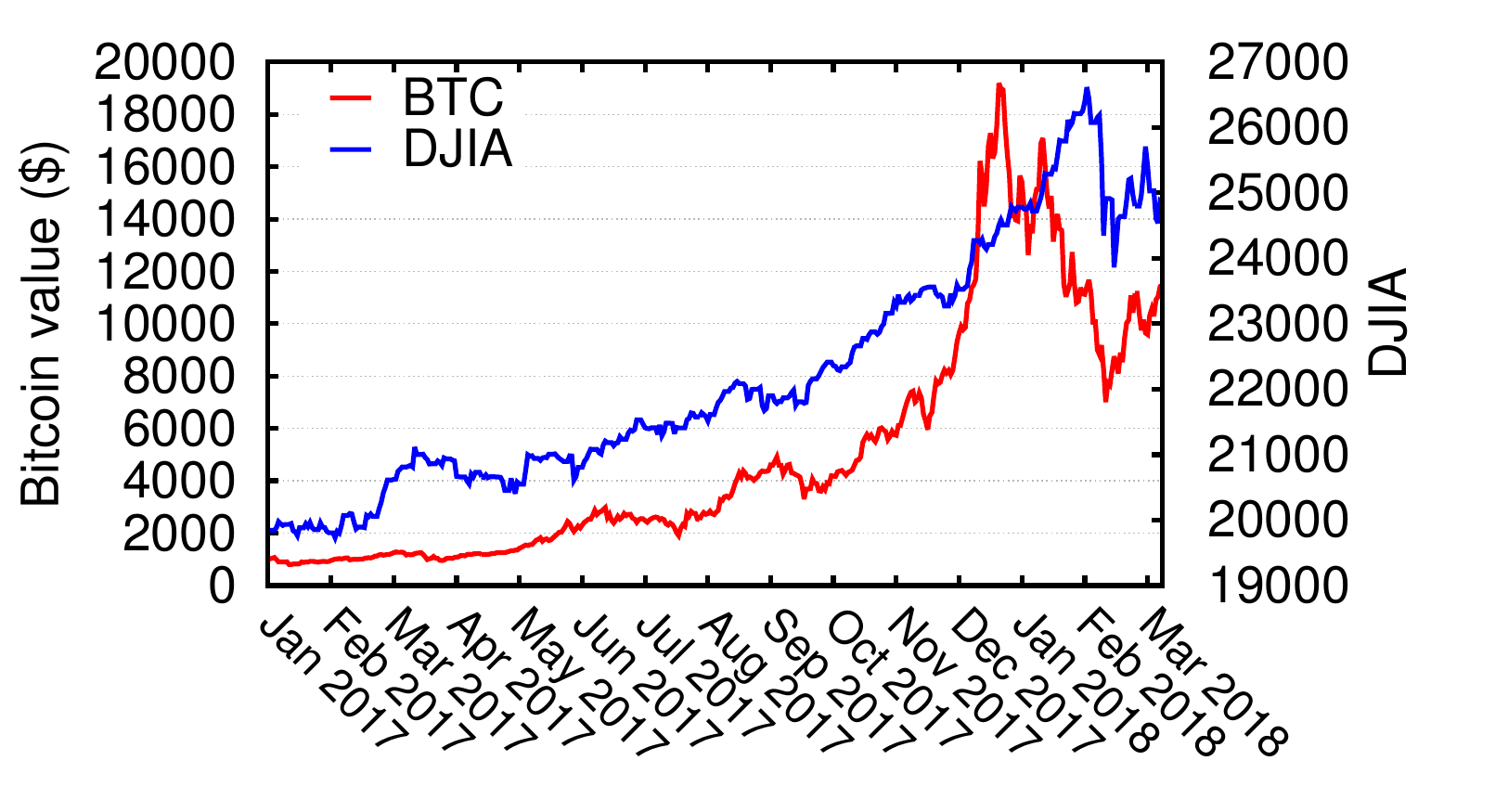}
		\caption{Values of Bitcoin and DJIA over the previous year of 2017. We see that the values tend to follow each other for most of the time.}
		\label{fig:DJ_vs_BTC17}
	\end{minipage}
\end{figure}

Additionally, since Dow Jones is closed on weekends we extrapolate using the volume of transactions occurred on Friday 
as values for Saturday and Sunday. Finally, to reduce the noise, we calculated the moving averages for each distribution 
and we plot the results in Figure~\ref{fig:volumeSince2014}.
As we can see, Bitcoin across almost the entire period of time, has an increasing number of transactions when the Dow Jones has a 
relatively stable volume. We see, thus that Bitcoin follows a totally independent pattern from the stock market of Dow Jones.
Indeed, we see that during the impressive increase of Bitcoin in 2016, Dow Jones remained stable for a while before slightly 
decreasing within the last months of the year.

Besides this non-correlation of the two distributions, what we see by focusing on the year of 2016 is the influence of 
the political incidents in the stock markets and how these do not seem to affect the Bitcoin. The year of 2016, and mostly 
the last half of it, offered a quite large number of political incidents worldwide~\cite{politics2016} such as (Brexit, 
the controversial win of D. Trump, the failure of Trans-Pacific 
Partnership, the North Korean nuclear tests, the Russian interference in U.S. Presidential elections, etc.). Apparently, these
incidents created a rather unstable environment for the worldwide economy and the stock markets, which can also be seen by 
the small plunge in the last months of 2016 before the first legislations of Donald Trump give a boost in the domestic economy 
of United States around January 2017. Contrary to that, we see a steady increase of the number of transactions in the Bitcoin 
throughout the entire year of 2016.

In our next experiment, we compared the value of Bitcoin in USD with the DJIA over the previous year. Our results, as can be seen in 
Figure~\ref{fig:DJ_vs_BTC17}, show that the value of Bitcoin and the value of DJIA tend to follow each other for most of the time. 
Indeed, for most of 2017 they seem to have an increasing trend. Although they diverge in early 2018 when Bitcoin takes a plunge, they 
quickly converge and follow each other again from mid January 2018. 
Although there are concerns that untraceability of Bitcoin provides a way to facilitate dark markets, we see that the value of Bitcoin 
follows Dow Jones and thus, seems to follow the rules of the open market. It seems that the vast majority of people use Bitcoin as an 
investment tool, much like the stock market and the rest of tools out there.

\section{Related Work}
\label{sec:related}
The last years cryptocurrencies and Bitcoin became a subject of research in both economics and computer science. 
Yermack~\cite{yermack} studies the ability of Bitcoin become an actual currency. He argues that money has 
three characteristics: (i) functions as medium of exchange, (ii) a unit of account, and (iii) a store of value. In addition, 
he argues that albeit Bitcoin can indeed be used as a medium of exchange, yet fails to satisfy the other two because of the 
high volatility in its price appearing to behave more like a speculative investment than a currency. Furthermore, Yermack 
correlates Bitcoin's daily exchange rate with the U.S. dollar, with the dollar's exchange 
rates against other prominent currencies, and also against gold.  
The result is a totally zero correlation, which in economics, means that Bitcoin has a risk nearly impossible to hedge 
for businesses and customers, thus rendering it more or less useless as a tool for risk management.

Daniel et al.~\cite{rich} use Bitcoin transactions to measure the dynamics taking place on the transaction network and analyze 
the flow of money, the temporal patterns and the wealth accumulation. Their results show that bitcoin wealth distribution is highly 
heterogeneous throughout the lifetime of the system, and it converges to a stable stretched exponential distribution in the trading phase.

In~\cite{dwyer2015economics}, authors discuss the rise of 24/7 trading on computerized markets in Bitcoin where there are no brokers or other agents. They note that the average monthly volatility of returns on Bitcoin is higher than for gold or a set of foreign currencies, however the lowest monthly volatilities for Bitcoin are less than the highest monthly volatilities for gold and the foreign currencies. Furthermore, although most people prefer to have their assets and liabilities denominated in the same currency (thus reducing their risk in terms of their own currency), this is not trivial in Bitcoin given the volatility of exchange rates. 

Papadopoulos et al. in~\cite{10.1007/978-3-030-30215-3_14} study the profitability of in-browser mining for the web publishers and the costs that this imposes on the user side in terms of system resources utilization, power consumption. Their results show that cryptomining can reach the profitability of advertising under specific circumstances, but users need to sustain a significant cost on their devices.

\section{Conclusion}
\label{sec:conlcusion}
The need of a substitute digital currency first expressed long time ago. The most successful 
attempt to design such a currency came 10 years ago with the birth of the first cryptocurrency, Bitcoin.
Bitcoin managed within this years to draw a lot of attention and skyrocket its market capitalization. Although, it is still far from being widely used as a currency, the 
wild up swings of its value, lures an increasing number of people to consider it as an asset to yield a trading profit.

In this study, we explore the investment landscape of the Bitcoin.
Our scope is to enhance the transparency on the different patterns followed in the worldwide, 
unregulated market of Bitcoin. To accomplish that, we obtain a large 5 year long 
snapshot of the blockchain and we extract the transactions. 
Finally, we compare our findings with the patterns followed in traditional 
stock market of Dow Jones.

\bibliographystyle{plain}
\balance
\bibliography{main}
\end{document}